\documentstyle[aps,twocolumn,prl,epsf,rotate]{revtex}

\tightenlines

\def\Ps56{{\cal P}_Z(s_{56})}
\def\A60{{\cal A}^{{\rm tree}}_6}
\def\d{{\rm d}}

\def\res#1#2{$\!\!#1\pm #2\!\!$}

\newcommand\xB{x_{\rm B}}
\newcommand\kT{k_\perp}
\newcommand\ycut{y_{\rm cut}}
\newcommand\dcut{d_{\rm cut}}
\newcommand\fcut{f_{\rm cut}}
\newcommand\as{{\alpha_s}}
\newcommand\asbar{\overline{\alpha}_s}

\newcommand\pdf{parton distribution function}
\newcommand\LO{LO}
\newcommand\NLO{NLO}

\newcommand\Qhs{Q_{\rm H.S.}}

\newcommand{\Nf}{{N_f}}

\newcommand{\beq}{\begin{equation}}
\newcommand{\eeq}{\end{equation}}
\newcommand{\beeq}{\begin{eqnarray}}
\newcommand{\eeeq}{\end{eqnarray}}
\newcommand\nn{\nonumber}
\newcommand\msbar{{\overline{{\rm MS}}}}
\newcommand{\KW}[1]{{\ttfamily\mdseries{#1}}}

\begin{document}

\begin{titlepage}
\begin{minipage}{15cm}

\begin{flushright}
hep-ph/0104315 \\
IPPP/01/18 \\
DCPT/01/36 \\
\end{flushright}

\par \vspace{10mm}
\begin{center}
{\Large \bf
Multi-jet cross sections in deep inelastic scattering at
next-to-leading order}
\end{center}
\par \vspace{2mm}
\begin{center}
{\bf Z. Nagy}\\
\vspace{3mm}
{\em Department of Physics, University of Durham, Durham DH1 3LE,
England}\\
{\it E-mail}: \KW{Zoltan.Nagy@durham.ac.uk}

\vspace{5mm}
{\bf Z. Tr\'ocs\'anyi}\footnote{Sz\'echenyi fellow of the
Hungarian Ministry of Education.}\\

\vspace{3mm}
{\em Department of Experimental Physics, KLTE,
H-4001 Debrecen P.O.Box 105, Hungary
\\
and Institute of Nuclear Research of the Hungarian Academy of Sciences}\\
{\it E-mail}: \KW{zoltan@zorro.atomki.hu}

\end{center}

\vspace{5mm}
\centerline{\bf Abstract}

\vspace{3mm}
We present the perturbative prediction for three-jet production cross
section in DIS at the \NLO\ accuracy. We study the dependence on the
renormalization and factorization scales of exclusive three-jet cross
section. The perturbative prediction for the three-jet differential
distribution as a function of the momentum transfer is compared to the
corresponding data obtained by the H1 collaboration at HERA.

\vspace{9cm}
~
\end{minipage}
\end{titlepage}
\newpage

\twocolumn[\hsize\textwidth\columnwidth\hsize\csname
@twocolumnfalse\endcsname

\title{Multi-jet cross sections in deep inelastic scattering at
next-to-leading order}
\author{Zolt\'an Nagy$^a$ and Zolt\'an Tr\'ocs\'anyi$^b$}
\address{
$^a$Department of Physics, University of Durham, Durham DH1 3LE, England
\\
$^b$Department of Experimental Physics, KLTE,
H-4010 Debrecen P.O.Box 5, Hungary
\\
and Institute of Nuclear Research of the Hungarian Academy of Sciences
}

\date{\today}
\maketitle

\begin{abstract}
We present the perturbative prediction for three-jet production cross
section in DIS at the \NLO\ accuracy. We study the dependence on the
renormalization and factorization scales of exclusive three-jet cross
section. The perturbative prediction for the three-jet differential
distribution as a function of the momentum transfer is compared to the
corresponding data obtained by the H1 collaboration at HERA.
\end{abstract}

]


Deep inelastic lepton-hadron scattering (DIS) has played a decisive role
in our understanding of the deep structure of matter. The latest version
of the experiment performed with colliding 27.5\,GeV electrons or
positrons and 820\,GeV protons at HERA yields increasingly precise data
so that not only fully inclusive measurements can be used to study the
physics of hadronic final states. In fact, the study of multi-jet
events and event shapes has become an important project at HERA
\cite{HERA}.

One of the important theoretical tools in the analysis of hadronic
final states is perturbative Quantum Chromodynamics (QCD).  In order to
make quantitative predictions in perturbative QCD, it is essential to
perform the computations (at least) at the next-to-leading order (NLO)
accuracy. In the case of DIS such computations have so far been
completed for one-jet inclusive and 2(+1 beam)-jet cross sections
\cite{DISNLO}. In this letter we compute the 3+1-jet exclusive cross
section, defined using the $\kT$ algorithm \cite{Catani:1992zp}, as a
function of the jet resolution parameter at the \NLO\ accuracy. We also
compute various differential distributions of the three-jet cross section,
defined using the inclusive $\kT$ algorithm \cite{Ellis:1993tq}, that can
be compared to recent data obtained by the H1 collaboration
\cite{H1_DIS01}.
With our Monte Carlo program one can compute the \NLO\ cross section
of any other infrared safe two- and three-jet quantity --- the
particular distributions shown here are given simply as illustration.

An immediate application of our computation is, of course, the analysis 
of HERA multi-jet data. Using our program, it is also possible to find
the higher order QCD correction to the forward jet cross section in the
low-$x$ regime in DIS, to which the O$(\alpha_s^2)$ computation effectively
gives only the leading-order prediction and large corrections may come
from higher orders \cite{forwardjet}. Our computation is also part of the
O$(\alpha_s^3)$ two-jet cross section.


In computing the \NLO\ corrections we use the dipole formalism of
Catani and Seymour \cite{Catani:1997vz} that we modify slightly in
order to have a better control on the numerical computation. The main
idea is to cut the phase space of the dipole subtraction terms as
introduced in Ref.~\cite{Nagy:1999bb}, the details of applying it to
the case of DIS will be given elsewhere.

The advantages of using the dipole method  are the followings: i) no
approximation is made; ii) the exact phase space factorization allows
full control over the efficient generation of the phase space; iii)
neither the use of color ordered subamplitudes, nor symmetrization, nor
partial fractioning of the matrix elements is required; iv) Lorentz
invariance is maintained, therefore, the switch between various frames
(e.g.\ laboratory and Breit frames) can be achieved by simply
transforming the momenta; v) the use of crossing functions is avoided;
vi) it can be implemented in an actual program in a fully process
independent way.


In order to ensure the correctness of our results we checked the
following points: (i) the correctness of all matrix elements and
phase-space generation; (ii) the subtraction term regularizes the real
correction in all possible soft and collinear limits; (iii) the same
quantity is subtracted from the real correction as added to the virtual
one.

We achieved full control on the matrix elements by utilizing the
well-tested matrix elements that we had used in our program 
{\tt DEBRECEN} \cite{DEBRECEN} for computing \NLO\ corrections to
three- and four-jet production in electron-positron annihilation
\cite{Nagy:1999bb}, with different particles crossed into the initial
state. In particular, we used the crossing symmetric tree-level
five-parton helicity amplitudes of Ref.~\cite{Nagy:1999bb} and the crossing
symmetric three- and four-parton tree and one-loop amplitudes of Bern et
al \cite{BDK}. We consider only virtual photon exchange, in which case
the necessary one-loop four-parton plus a gauge boson matrix elements
were also derived by Glover et al \cite{CGM}. Both sets of matrix
elements were tested in $e^+e^-\to$ four jets \NLO\ cross section
computations and full agreement among the theoretical descriptions
\cite{NLO4jet} and very good description of experimental data
\cite{NTQCD98} were found. 

We have checked numerically that in all soft and collinear regions the
difference of the real and subtraction terms contain only integrable
square-root singularities. Furthermore, we have also checked that our
results are independent of the parameter that controls the volume of
the cut dipole phase space, which ensures that indeed the same quantity
has been subtracted from the real correction as added to the virtual
one.

Finally, to have a further check of the computation, we utilized that
the dipole method allows for the construction of a process independent
programming of QCD jet cross sections at the \NLO\ accuracy.  We use the
same program structure, with trivial modifications, to compute dijet
and three-jet cross sections.  Consequently, using the correct matrix
elements and checking that our dijet cross sections are correct, we
obtain the correctness of the three-jet result automatically.

In order to check the strucure of our program we compared our \NLO\
two-jet predictions to those of two other existing programs
\cite{DISENT,DISASTER}.  In Ref.~\cite{DISASTER} a comparison of the
these programs was presented using the \NLO\ two-jet results for the
modified JADE clustering scheme. In Table~I. we recall the numbers
presented there for one particular set of \pdf s together with the
corresponding result of our computation. (We refer to
Ref.~\cite{DISASTER} for the precise meaning of these numbers.) We find
complete agreement with the predictions of {\tt DISASTER++}. Our
results also agree with the predictions of {\tt DISENT} with slight
tendency to lower values, but within statistical errors, from 
{\tt DISENT} for small values of $x_B$ and the inelasticity $y$ (first
two bins). A more complete comparison will be presented elsewhere.

\vbox{
\begin{table}             
\caption{Comparison of the \NLO\ two-jet cross sections defined using the
modified JADE algorithm and for MRSD$'_-$ parton densities obtained with
three partonic Monte Carlo programs {\tt DISENT},
{\tt DISASTER++} and {\tt NLOJET++} (this work).}
\begin{tabular}{cccc}
bin &  {\tt DISENT}  &{\tt DISATER++} &{\tt NLOJET++}   \\   
\tableline
 1  &\res{578.4}{7.1}&\res{585.0}{2.6}&\res{585.5}{2.0} \\
 2  &\res{361.1}{3.5}&\res{364.8}{1.5}&\res{364.8}{1.5} \\
 3  &\res{120.1}{0.9}&\res{119.1}{1.7}&\res{122.9}{0.5} \\
 4  &\res{95.4}{0.87}&\res{98.1}{1.11}&\res{97.7}{0.6} \\
 5  &\res{54.9}{0.40}&\res{55.3}{0.46}&\res{55.7}{0.4} \\
 6  &\res{17.3}{0.13}&\res{17.5}{0.06}&\res{17.5}{0.1} \\
 7  &\res{12.3}{0.15}&\res{12.1}{0.50}&\res{12.4}{0.07} \\
 8  &\res{8.52}{0.08}&\res{8.61}{0.12}&\res{8.61}{0.07} \\
 9  &\res{2.63}{0.02}&\res{2.65}{0.03}&\res{2.63}{0.02} \\
\end{tabular}                                               
\end{table}
}


In the case of DIS the observables may be defined either in the
laboratory frame, in the hadronic center of mass frame (virtual boson
plus proton rest frame), or in the Breit frame, which is characterized
by purely spacelike virtual boson momentum. In order to easily apply
detector cuts, our program generates events in the laboratory frame. 
Due to the Lorentz invariance of the dipole method, we can freely
transform the momenta to the other frames to compute the jet functions.
Jet algorithms can be defined in any of these frames. The $\kT$ scheme
we employ here is implemented also in the Breit frame \cite{Catani:1992zp}.
The full algorithm is a two-step procedure, the first one is a
preclustering of hadrons characterized by a pre-chosen stopping parameter
$\dcut$. For clustering the partons, we used the covariant
$E$-scheme. The result of the first step is the beam jet plus hard
final-state jets. The second step is the resolution of the hard
final-state jets into sub-jets characterized by a parameter $\ycut$.
We scaled the stopping parameter with the momentum transfer squared
$Q^2$ as $\dcut = \fcut Q^2$, and we chose $\ycut = 1$, when the second
step is not carried out. Thus our three-jet cross section is a function
of one parameter, $\fcut$.

Once the phase-space integrations are carried out, we write the 
three-jet cross section (at a given $\fcut$) at \NLO\ accuracy in the
following form:
\beeq
\label{sig3}
&&
\sigma_{\rm 3jet}(\fcut)
= \sum_a\!\int\!\d \eta\, f_a(\eta,\mu_F) 
\Big[ \asbar(\mu_R)^2 B_a(\eta,\fcut)
\nn \\ &&\quad
+ \asbar(\mu_R)^3
\Big(B^{(R)}_a(\eta,\fcut, \mu_R)
   + B^{(F)}_a(\eta,\fcut, \mu_F)
\nn \\ &&\qquad\qquad\qquad
   + C_a(\eta,\fcut)
\Big)\Big]\:.
\eeeq
In Eq.~(\ref{sig3}) $\asbar = \as/2\pi$, $f_a(\eta,\mu_F)$ is the \pdf\
for parton type $a$ at momentum fraction $\eta$ and factorization scale
$\mu_F$.  $B_a$ gives the Born contribution, $B^{(R)}_a$, $B^{(F)}_a$,
$C_a$ are the correction functions.  In the higher order
correction we separated the dependence on the renormalization scale
$\mu_R = x_R \Qhs$ and factorization scale $\mu_F = x_F \Qhs$, where
$\Qhs$ is the hard scattering scale. $\Qhs$ is usually set event by
event, therefore, the scale dependences can only be written in the
factorized form if we consider differential cross section in $\Qhs$.
The $B^{(R)}_a(\eta,\fcut,\mu_F)$ function is obtained by multiplying
the Born squared matrix element with $\beta_0 \ln x_R^2$, where
$\beta_0 = \frac13\,(11C_A - 2\Nf)$, and integrating over the phase
space.  Apart from trivial factors the $B^{(F)}_a(\eta,\fcut,\mu_F)$
function is obtained by convoluting the Altarelli-Parisi splitting
functions with the integral of the colour-correlated Born squared
matrix elements (see Eqs.~(8.39) and (8.41) in
Ref.~\cite{Catani:1997vz}). The $C_a(\eta,\fcut)$ correction function
requires much more computer time to compute.

Eq.~(\ref{sig3}) shows that using the dipole method one may either
compute the full cross section at the \NLO\ accuracy including the
convolution with the \pdf s, or simply the parton level functions
$B_a(\eta,\fcut)$, $B^{(F)}_a(\eta,\fcut)$ and $C_a(\eta,\fcut)$,
which can then be convoluted with the parton densities after the Monte
Carlo integration. The latter procedure is the proper one if we are
interested in an $\as$ measurement from DIS data (to avoid the
reculculation of the Monte Carlo integrals after each step of the
fitting iteration). However, it requires at least triple differential
binning (one for $\fcut$, one for the momentum fraction $\eta$ and one for
the hard scattering scale $\Qhs$) for the nine partonic functions
(three for each of the three incoming flavor type), which is beyond the
scope of a letter. 

When presenting numerical results of three-jet cross sections in this
paper we use the kinematic region used by the H1 collaboration
\cite{H1_DIS01}. For the basic DIS kinematic variables $Q^2$, $\xB$
and $y = Q^2/(s\,\xB)$ we require
\beeq
&&
5\,{\rm GeV}^2 < Q^2 < 5000\,{\rm GeV}^2\,,
\nn \\ &&
0 < x_{\rm Bj} < 1\,,\quad
0.2 < y < 0.6\:.
\eeeq
Furthermore, we restrict the (pseudo)rapidity-range in the laboratory
frame and the minimum transverse energy of the jets in the Breit frame as
\beq
-1 < \eta_{\rm jet, lab} < 2.5\,,\qquad E_T > 5\,{\rm GeV}\:.
\eeq
We choose the average transverse momentum of the jets,
\beq
\label{eq:Qhs}
\Qhs = \frac13 \sum_j E_T^B(j)\:
\eeq
as hard scattering scale. We also studied the other usual choice, when
$\Qhs^2$ is the negative of the momentum transfer carried by the
virtual boson ($Q^2$), but have not found significant differences.

\begin{figure}
\epsfxsize=8cm \epsfbox{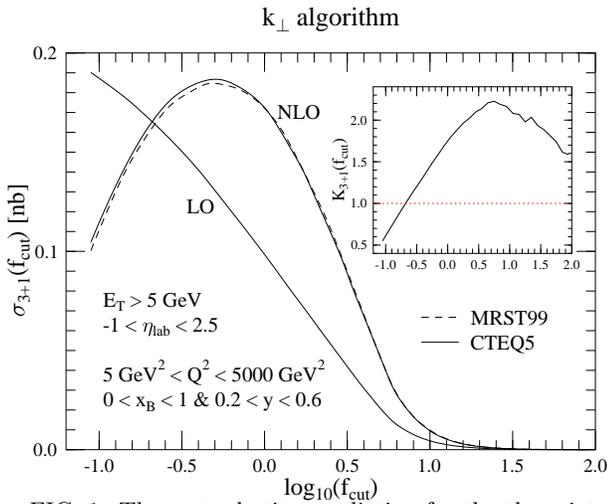}
\caption{The perturbative prediction for the three-jet cross section
$\sigma_{\rm 3jet}(\fcut)$ as a function of $\fcut$.  The solid lines
were obtained using the CTEQ5 \pdf s and the dashed lines with the
MRST99 set.}
\label{fig:sig-fcut}
\end{figure}

In Fig.~\ref{fig:sig-fcut}, we plotted the cross section convoluted with
the CTEQ5M1 \pdf s \cite{CTEQ5} and using the two-loop formula for the
strong coupling, 
\beq
\label{twoloopas}
\asbar(\mu) = \frac{\asbar(M_Z)}{w(\mu,M_Z)}
\left(
1-\asbar(M_Z)\,\frac{\beta_1}{\beta_0}
\frac{\ln(w(\mu,M_Z))}{w(\mu,M_Z)}
\right)\:,
\eeq
where $w(q,q_0) = 1 - \beta_0\,\asbar(q_0)\,\ln(q_0/q)$,
$\asbar(M_Z) = 0.118/2\pi$
and $\beta_1 = \frac13\,(17C_A^2 - 6 C_F T_R \Nf - 10 C_A T_R \Nf)$,
with $\Nf = 5$ flavors. For the leading order results we used the
CTEQ5L distributions and the one-loop $\as$ ($\asbar(M_Z) = 0.127/2\pi$
and $\beta_1 = 0$ in Eq.~(\ref{twoloopas})).  We used running $\msbar$
electromagnetic coupling at the scale of the virtual photon momentum
squared.  In order to show the weak dependence on the \pdf s, we
plotted the perturbative predictions also using the MRST99 \pdf s
\cite{MRST99} (dashed lines). There is no \LO\ fit in the MRST99 set,
therefore, in this latter case the LO curve is missing.

The inset in Fig.~\ref{fig:sig-fcut} shows the $K$ factor (ratio of the
three-jet cross section at the \NLO\ to that at the \LO\ accuracy),
indicating the relative size of the correction. The $K$ factor exhibits
a clear maximum, where the correction is more than 100\,\% and becomes
smaller than one, implying negative correction, for $\fcut < 0.2$. The
position and height of the maximum as well as that of the point $K=1$
depends on the lower limit on the $Q^2$ range. For instance, increasing
the lower limit from $5\,{\rm GeV}^2$ to $100\,{\rm GeV}^2$, the point
$K=1$ moves to about $\fcut = 0.03$. Negative \NLO\ correction
indicates that the resummation of large logarithms is necessary in
order to obtain a reliable prediction for smaller values of the
stopping parameter.  

\begin{figure}
\epsfxsize=8cm \epsfbox{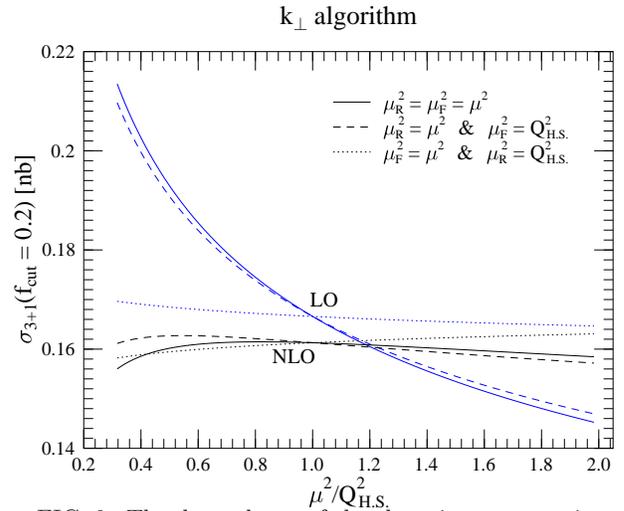}
\caption{The dependence of the three-jet cross section $\sigma_{\rm
3jet}$ at the fixed value of $\fcut = 0.2$ on the renormalization and
factorization scales. }
\label{fig:mudep}
\end{figure}

In Fig.~\ref{fig:mudep} we study the scale dependences of the three-jet
cross section at a fixed value of the resolution parameter $\fcut = 0.2$. 
The strong dependence on the renormalization scale observed at \LO\ is
significantly reduced. The factorization scale dependence is already not
significant at \LO\ and does not change much. Setting the two scales
equal, $\mu_R = \mu_F = \mu$, we can observe a wide plato peaking near
$\mu = \Qhs$.

\begin{figure}
\epsfxsize=8cm \epsfbox{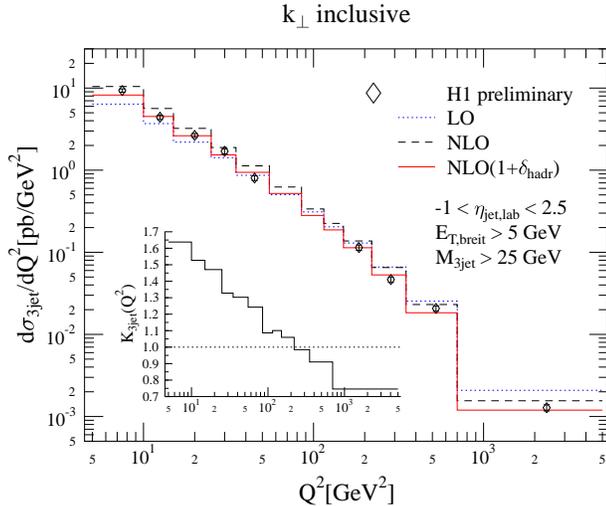}
\caption{The differential three-jet cross section 
for the $5 < Q^2 < 5000\,{\rm GeV}^2$ region compared to experimental
data measured by the H1 collaboration.  The \LO\ prediction (dashed
line) was obtained with the CTEQ5L \pdf s and the \NLO\ prediction
(solid line) was obtained with the CTEQ5M1 \pdf s.}
\label{fig:sig-Q2}
\end{figure}

The first measurement of the three-jet cross sections in DIS was
published by the H1 collaboration in Ref.~\cite{H1_DIS01}. They used
the inclusive $\kT$ algorithm to define the jets (the precise definition
is given in Ref.~\cite{Adloff:1999ni}), selected three-jet
events and plotted differential distributions of certain kinematical
variables. We computed the same distributions at the \LO\ and \NLO\
accuracy and compared the predictions, corrected to hadron level (the
correction factors are about 0.8), to the published data. In
Fig.~\ref{fig:sig-Q2} we present the $\d\sigma_{\rm 3jet}/\d Q^2$
differential distribution.  We observe that the \LO\ prediction has a
different shape than the data: too low for small values of $Q^2$ and
too high at high values of $Q^2$. The radiative corrections bring
theory and experiment much closer: the \NLO\ prediction gives a
remarkably good description of the corrected experimental data. Similar
conclusions can be drawn from the distibutions of the other kinematical
variables studied in Ref.~\cite{H1_DIS01}, but not shown here.  


In this letter we presented a \NLO\ computation of the three-jet rate 
defined with the $\kT$ clustering algorithm in DIS.  Our results were
obtained using a partonic Monte Carlo program that is suitable for
implementing any detector cuts. We found that the $K$ factor is very
sensitive to the allowed kinematic region. We demonstrated that the
\NLO\ corrections reduce the scale dependence significantly. The \NLO\
prediction is weakly dependent on the \pdf s. The \NLO\ perturbative
prediction gives a remarkably good description of data measured at HERA
and corrected to parton level.  The same program can
be used for computing the QCD radiative corrections to
the (differential) cross section of any kind of two-, or three-jet
cross section or event-shape distribution in DIS.  We compared the
two-jet rates obtained by our program to previous results and
found agreement.

\medskip
We thank M.~Wobisch for communicating the H1 data \cite{H1_DIS01} to us.
This work was supported in part by the EU Fourth Framework Programme
``Training and Mobility of Researchers'', Network ``QCD and particle
structure'',
contract FMRX-CT98-0194 (DG 12 - MIHT), the EU Fifth Framework
Programme `Improving Human Potential', Research Training Network
`Particle Physics Phenomenology at High Energy Colliders', contract
HPRN-CT-2000-00149 as well as by the Hungarian Scientific Research Fund
grant OTKA T-025482.

\end{document}